\documentclass[letter]{aa}

\usepackage{graphicx}
\usepackage{txfonts}
\usepackage{longtable}
\usepackage{float}

\def\m31{{M~31}}

\begin{document} 

\title{A remarkable recurrent nova in \m31: The 2010 eruption recovered and evidence of a six-month period.}

\author{M.~Henze\inst{1}\thanks{Corresponding authors}
\and M.~J. Darnley\inst{2\,\star}
\and F.~Kabashima\inst{3}
\and K.~Nishiyama\inst{3}
\and K.~Itagaki\inst{4}
\and X.~Gao\inst{5}
}

\institute{Institut de Ci\`encies de l'Espai (CSIC-IEEC), Campus UAB, C/Can Magrans s/n, 08193 Cerdanyola del Valles, Spain \\
  \email{henze@ice.cat}
  \and Astrophysics Research Institute, Liverpool John Moores University, IC2 Liverpool Science Park, Liverpool, L3 5RF, UK \\
  \email{M.J.Darnley@ljmu.ac.uk}
  \and Miyaki-Argenteus Observatory, Miyaki, Saga-ken, Japan
  \and Itagaki Astronomical Observatory, Teppo, J990-2492 Yamagata, Japan
  \and Xingming Observatory, Mt. Nanshan, Urumqi, Xinjiang, China
}
\date{Received 2015 August ? / Accepted ?}

\abstract{The Andromeda Galaxy recurrent nova M31N~2008-12a has been caught in eruption nine times. Six observed eruptions in the seven years from 2008 to 2014 suggested a duty cycle of $\sim1$ year, which makes this the most rapidly recurring system known and the leading single-degenerate Type Ia Supernova progenitor candidate; but no 2010 eruption has been found so far. Here we present evidence supporting the recovery of the 2010 eruption, based on archival images taken at and around the time. We detect the 2010 eruption in a pair of images at 2010 Nov 20.52 UT, with a magnitude of $m_{R}=17.84\pm0.19$. The sequence of seven eruptions shows significant indications of a duty cycle slightly shorter than one year, which makes successive eruptions occur progressively earlier in the year. We compared three archival X-ray detections with the well observed multi-wavelength light curve of the 2014 eruption to accurately constrain the time of their optical peaks. The results imply that M31N~2008-12a might have in fact a recurrence period of $\sim6$ months ($175\pm11$\,days), making it even more exceptional. If this is the case, then we predict that soon two eruptions per year will be observable. Furthermore, we predict the next eruption will occur around late Sep 2015. We encourage additional observations.}


\titlerunning{M31N~2008-12a: 2010 eruption recovered and evidence of a 6-months period}

\maketitle

\section{Introduction}\label{intro}

Classical and recurrent novae are a subclass of the cataclysmic variables that display luminous eruptions driven by a thermonuclear runaway on the surface of an accreting white dwarf \citep[WD; see][for recent reviews]{2008clno.book.....B,2010AN....331..160B,2014ASPC..490.....W}.  By definition, a recurrent nova (RN) is any classical nova (CN) that has been observed in eruption more than once.  Such a classification is limited by observational selection effects, certainly at the upper end, with observed recurrence times between 1--100 years.  \citet{2014ApJ...793..136K} predicted the lower limit of recurrence times to be $\sim2$\,months (for a 1.38\,$M_{\sun}$\ WD with a mass accretion rate of $3.6\times10^{-7}$\,$M_{\sun}$\,yr$^{-1}$).

Since the time of Edwin \citet{1929ApJ....69..103H}, novae have been readily observed extragalactically, particularly in \m31, which with an annual nova rate of $65^{+16}_{-15}$ eruptions \citep{2006MNRAS.369..257D}, and almost 1\,000 nova eruptions discovered to date \citep[also see the on-line database\footnote{\url{http://www.mpe.mpg.de/~m31novae/opt/m31/index.php}}]{2007A&A...465..375P} is the prime laboratory in which to study these systems.  \citet{2015ApJS..216...34S} has recently published an astrometric catalogue of 16 \m31 RNe, whereas \citet{2014ApJS..213...10W,2015Wil} has released a catalogue of 11 \m31 nova progenitor systems.  Both these studies indicate that the contribution of RNe to the overall nova population may be significantly larger than originally thought, in line with Galactic results from \citet{2014ApJ...788..164P}.

The remarkable recurrent nova \object{M31N 2008-12a} exhibits the shortest eruption duty cycle of any known nova, just 1~year \citep{2014A&A...563L...9D,2015RN}. Its short recurrence time is driven by a high mass WD and high mass accretion rate \citep[][]{2014A&A...563L...8H,2015Hen,2014ApJ...793..136K,2015arXiv150605364K}.  The first optical eruption was detected in 2008 Dec, with subsequent optical detections in 2009 Dec, 2011 Oct, 2012 Oct, 2013 Nov, and 2014 Oct (see Table~\ref{eruption_history} for details).  Serendipitous X-ray detections, later associated with earlier eruptions, were seen in 1992 Feb, 1993 Jan, and 2001 Sep \citep{2014A&A...563L...8H,2014ApJ...786...61T}.

Initially, the `missing' 2010 eruption was not seen as troublesome until the full picture of the eruption history started to emerge after the 2013 eruption.  At the time, only the 2008 eruption was known \citep[the 2009 eruption was first announced by][]{2014ApJ...786...61T} and the system had yet to be spectroscopically confirmed.  But with the growing wealth of eruptions and data, the hunt has been on to uncover evidence of an initially missed 2010 eruption. Based on, for example, data from the Palomar Transient Factory \citep[see their Figure~4]{2012ApJ...752..133C}, the largest windows of time in which the 2010 eruption could have occurred were Oct 18--26 or Nov 19--26.

\begin{table*}
\caption{List of observed eruptions of \object{M31N 2008-12a}.\label{eruption_history}}
\begin{center}
\begin{tabular}{lllll}
\hline
\hline
$t_{\mathrm{max,\,optical}}$\tablefootmark{a} & $t_{\mathrm{max,\,X-ray}}$\tablefootmark{b} & Days since & Source & References \\
(UT) & (UT) & last eruption\tablefootmark{c} \\
\hline
(1992 Jan 29) & 1992 Feb 05 &  & X-ray ({\it ROSAT}) & 1 \\
\hline
(1993 Jan 04) & 1993 Jan 11 & 341 & X-ray ({\it ROSAT}) & 1 \\
\hline
(2001 Aug 26) & 2001 Sep 08 & & X-ray ({\it Chandra}) & 2 \\
\hline
2008 Dec 26 & & & Optical & 3 \\
\hline
2009 Dec 03 & & 342 & Optical (Palomar Transient Factory) & 4 \\
\hline
2010 Nov 20 & & 353 & Optical & 5 \\
\hline
2011 Oct 23.49 & & 337.5 & Optical & 4, 6, 7, 8 \\
\hline
2012 Oct 19.72 & $<2012$ Nov 06.45 & 362.2 & Optical; X-ray ({\it Swift}) & 8, 9, 10, 11 \\
\hline
2013 Nov 28.60 & 2013 Dec $05.9\pm0.2$ & 405.1 & Optical (iPTF); X-ray ({\it Swift}) & 4, 8, 11, 12, 13 \\
\hline
2014 Oct $03.7\pm0.1$ & 2014 Oct 13.6 & 309.1 & Optical (Liverpool Telescope); X-ray ({\it Swift}) & 8 \\
\hline
\end{tabular}
\end{center}
\tablefoot{
Updated version of Table~1 from \citet{2014ApJ...786...61T}.  \tablefoottext{a}{Time of the optical peak with, those in parentheses are extrapolated from the X-ray data (see Sect.\,\ref{sec:disc_stat}).}\tablefoottext{b}{Time of the X-ray peak.}\tablefoottext{c}{Time since last eruption only quoted when consecutive detections occurred in consecutive years.}}
\tablebib{
(1)~\citet{1995ApJ...445L.125W}, (2)~\citet{2004ApJ...609..735W}, (3)~\citet{2008Nis}, (4)~\citet{2014ApJ...786...61T}, (5)~This Letter, (6)~\citet{2011Kor}, (7)~\citet{2011ATel.3725....1B}, (8)~\citet{2015RN}, (9)~\citet{2012Nis}, (10)~\citet{2012ATel.4503....1S}, (11)~\citet{2014A&A...563L...8H}, (12)~\citet{2013ATel.5607....1T}, (13)~\citet{2014A&A...563L...9D}, (14)~\citet{2015Hen}.
}
\end{table*}

In this Letter we present our recovery of the 2010 eruption of \object{M31N 2008-12a}.  In Section~\ref{obs} we describe the archival observations and their analysis.  In Section~\ref{disc} we present our results and discuss their impact on the recurrence time scale. We draw our conclusions in Section~\ref{conc}.

\section{Observations and data analysis}\label{obs}

We analysed archival optical data obtained with the following telescopes: (a) a Meade 200R 40~cm f/9.8 reflector, plus SBIG STL1001E camera, at \textit{Miyaki-Argenteus} observatory, Japan (observers: F.~Kabashima and K.~Nishiyama); (b) a 50~cm f/6 telescope, with BITRAN BN-52E(KAF-1001E) camera, at \textit{Itagaki} Astronomical Observatory, Japan (observer: K.~Itagaki); and (c) a 35~cm f/6.9 Celestron C14 Schmidt-Cassegrain telescope at \textit{Xingming} observatory, China (observer: X.~Gao). All observations were unfiltered and their dates are given in Table~\ref{obslog}.

All images were reduced and calibrated in a homogeneous way, using the AstrOmatic\footnote{\url{http://www.astromatic.net/}} software packages, \textit{SExtractor} \citep[][]{1996A&AS..117..393B} for source extraction, \textit{SWarp} \citep[][]{2002ASPC..281..228B} for image stacking, and \textit{SCAMP} \citep[][]{2006ASPC..351..112B} for astrometric and photometric calibration. We optimised the image reduction procedures to correct for the background light of \m31 and specific detection thresholds were used to create clean source catalogues. The astrometric solutions were computed in SCAMP, using as a reference system the \m31 part of the Local Group Galaxy Survey \citep[LGGS;][]{2006AJ....131.2478M}.

Aperture photometry was performed on all these images using SExtractor (v2.19.5), point-spread function (PSF) fitting was also performed by the Starlink {\it photom} package \citep[v1.12-2;][]{1982QJRAS..23..485D} for comparative purposes.  Photometric calibration was achieved using three LGGS stars  common to all the images \citep[stars \#12, \#14, and \#15, see][their Table~2]{2015RN}.  Although these observations were all unfiltered, the data for the secondary standards is well matched to the $R$-band, with typical calibration accuracy $<2\%$. 

No source was detected at the position of \object{M31N 2008-12a} in eight of the ten observations, see Table~\ref{obslog}.  The faintest nearby resolved object in all of these images, \object{J004528.55+415451.7} \citep[star \#11,][]{2015RN}, was detected with a S/N of $\sim10$ in all images. This object has a similar brightness to the peak of \object{M31N 2008-12a}.  The average photometry of this object yielded $m_{R}=17.880\pm0.049$, consistent with the LGGS value of $m_{R}=17.876$.  In Table~\ref{obslog} we record the $3\sigma$ limiting magnitudes of these eight observations. All ten observations were capable of detecting the nova at least 0.5~magnitudes below peak light.

A source was found at the position of \object{M31N 2008-12a} in a pair of observations taken on 2010 Nov 20.  In the first observation, taken at Nov 20.503 UT, a source was detected with S/N=4.2. Photometry of this source yielded $m_{R}=17.63\pm0.26$.  In the second observation, 0.036~days later, the source appeared to have faded to $18.24\pm0.42$ (S/N=2.6).  The stacking of  these two observations gives $m_{R}=17.84\pm0.19$ (S/N=5.7).

\begin{figure*}
\includegraphics[width=0.33\textwidth]{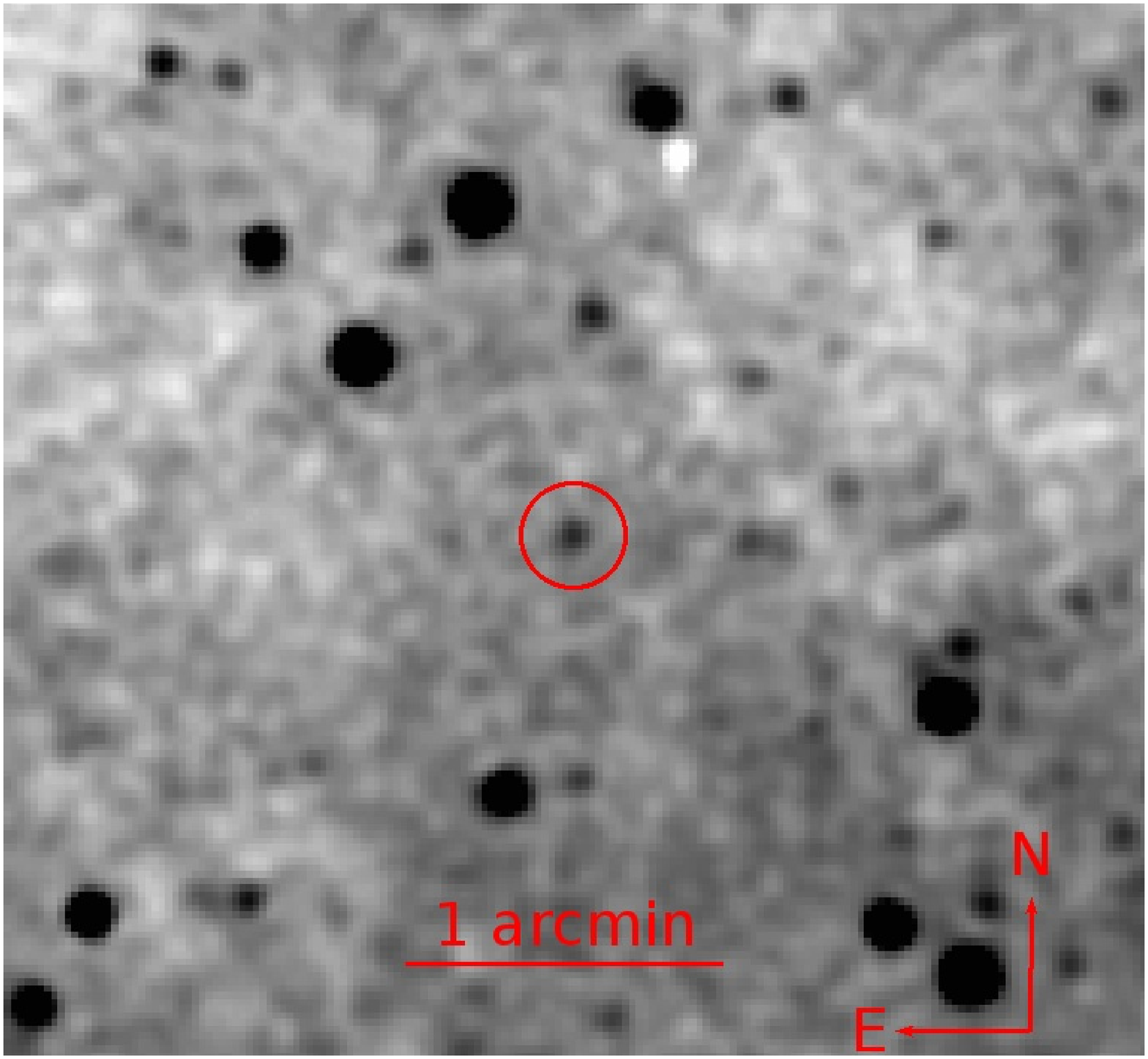}\hspace{0.005\textwidth}
\includegraphics[width=0.33\textwidth]{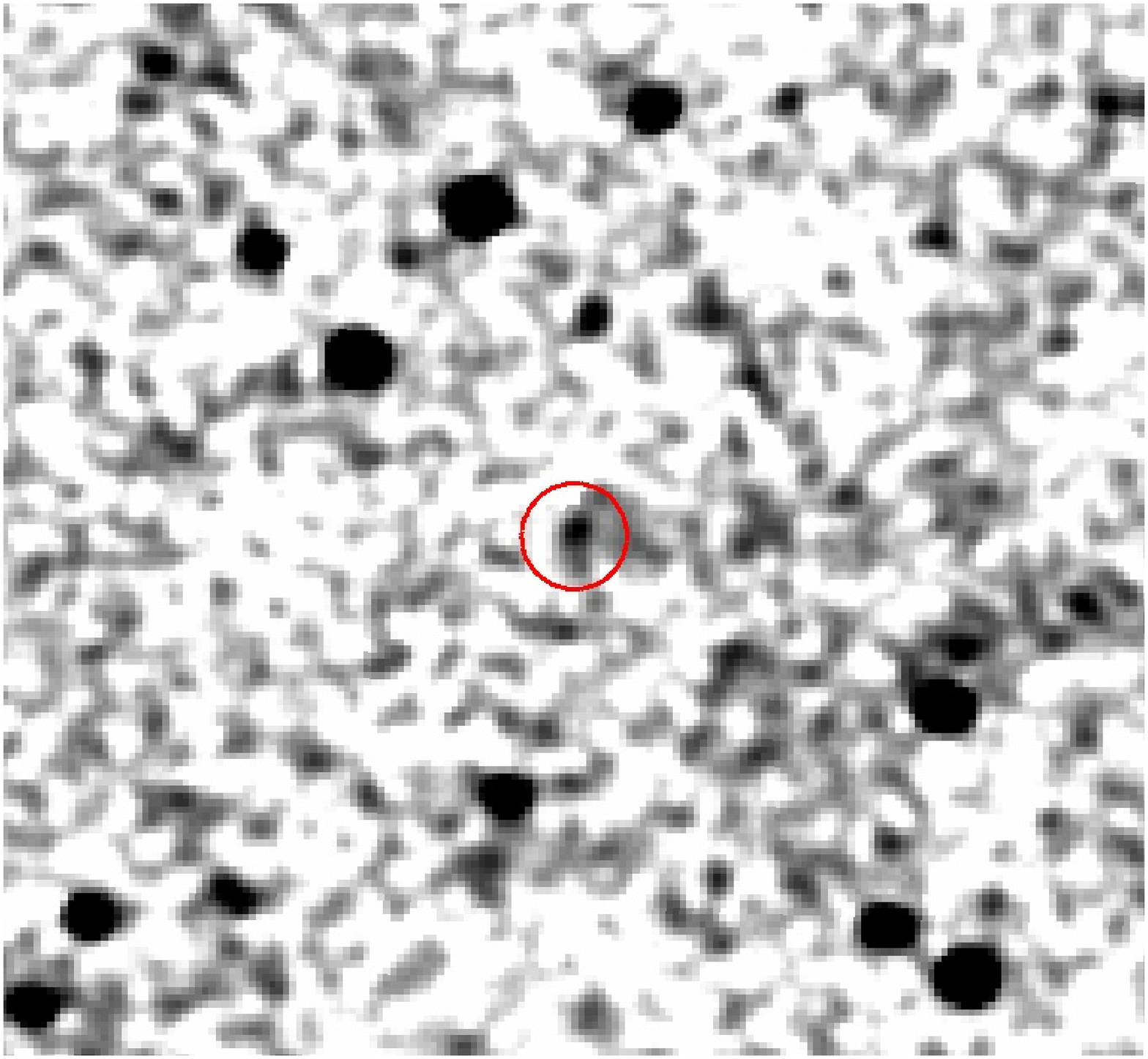}\hspace{0.005\textwidth}
\includegraphics[width=0.33\textwidth]{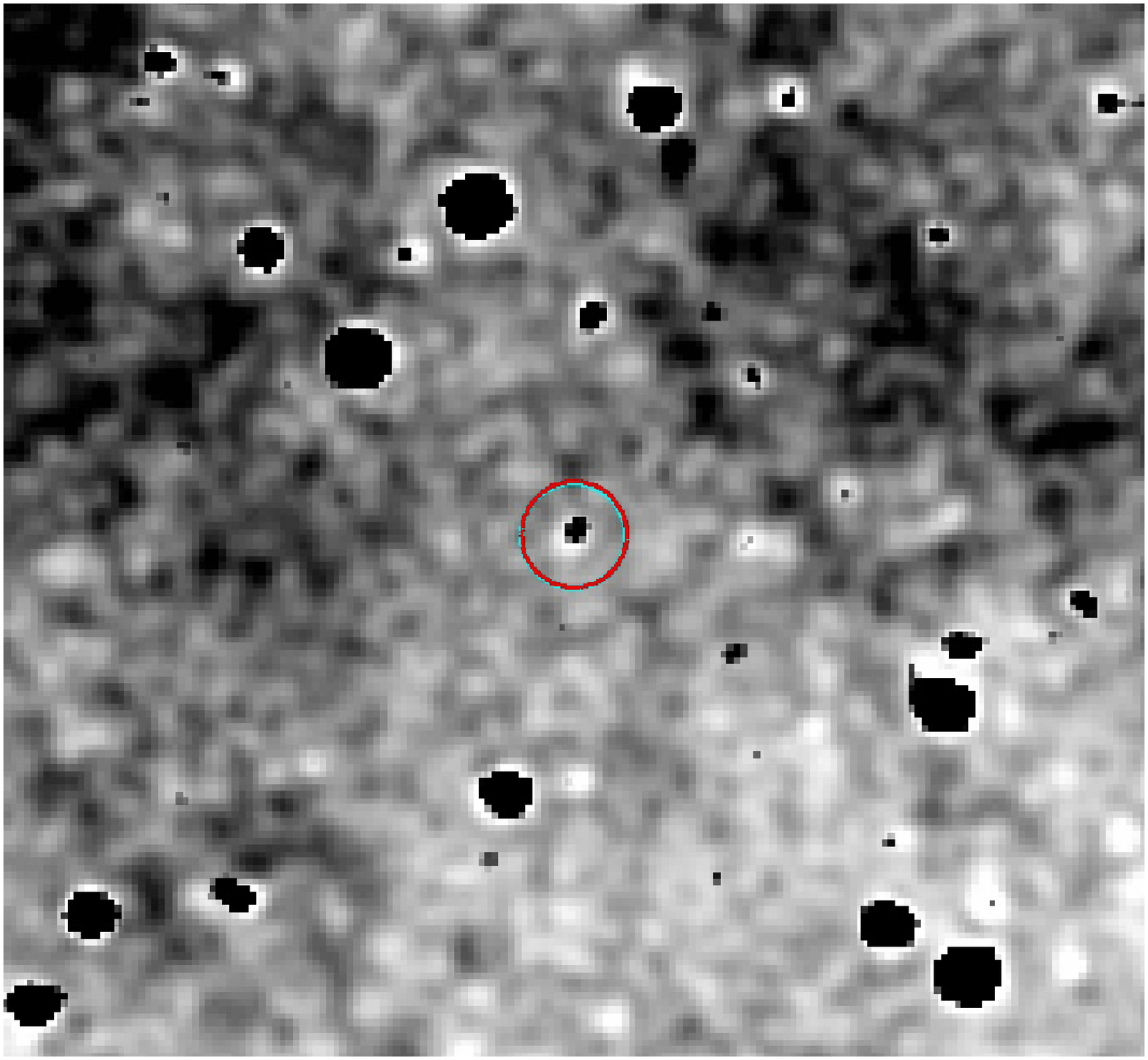}
\caption{Comparison of the \object{M31N 2008-12a} detections in 2010 and 2012. \textit{Left:} Stacked, smoothed image of the 2012 detection based on Miyaki-Argenteus data. \textit{Middle:} Stacked, background-subtracted, and smoothed image combining the two detection images in Table\,\ref{obslog}. \textit{Right:} Overlay of the smoothed 2010 and 2012 stacks, with the 2012 image inverted and in the background and the 2010 image having white turned transparent.}
\label{fig:blink}
\end{figure*}

In Fig.\,\ref{fig:blink} we show the positional agreement between these detections and the 2012 eruption of \object{M31N 2008-12a}, discovered with the same telescope. The comparison plot (right panel) has been created by inverting the 2012 image and overlaying it with the 2010 image, which had its white colour channel turned transparent. We also changed the colour cuts of the 2010 image to improve the contrast. Residual deviations between the two WCS solutions were corrected by eye to have the neighbouring stars match as closely as possible. The plot clearly shows that the positions of the two detection are consistent within their (similar) PSFs. Both are also consistent with the position of \object{M31N 2008-12a} obtained during the 2013 eruption: RA = $0^{\mathrm{h}}45^{\mathrm{m}}28^{\mathrm{s}}\!.81$, Dec = +$41\degr54\arcmin9\farcs9$ \citep[J2000;][]{2014A&A...563L...9D}.

\begin{table}
\caption{Observation log}\label{obslog}
\begin{center}
\begin{tabular}{lll}
\hline
\hline
Date (UT) & Observatory & Photometry\\
\hline
2010 Nov 11.548 & Miyaki-Argenteus & $>19.0$ \\
2010 Nov 12.515 & Xingming & $>18.9$ \\
2010 Nov 13.526 & Xingming & $>19.1$ \\
2010 Nov 14.550 & Xingming & $>19.0$ \\
2010 Nov 20.503 & Miyaki-Argenteus & $17.63\pm0.26$ \\
2010 Nov 20.539 & Miyaki-Argenteus & $18.24\pm0.42$ \\
2010 Nov 21.385 & Itagaki & $>18.8$ \\
2010 Nov 24.517 & Miyaki-Argenteus & $>19.5$ \\
2010 Nov 25.525 & Xingming & $>19.4$ \\
2010 Nov 26.478 & Miyaki-Argenteus & $>19.8$ \\
\hline
\end{tabular}
\end{center}
\end{table}

\section{Discussion}\label{disc}

\subsection{The 2010 eruption}
\label{sec:disc_2010}

The missing 2010 eruption of \object{M31N 2008-12a} had been the one downside in the unfolding story of this remarkable recurrent nova.  The 1992, 1993, and 2001 eruptions uncovered in archival X-ray observations by \citet{2014A&A...563L...8H} and \citet{2014ApJ...786...61T} strongly indicated that yearly eruptions had been ongoing for a number of decades.  The discovery of a possible vast remnant around the system by \citet{2015RN} provides a tantalising hint at an eruption history over a significantly longer time-scale.

The pair of clear detections of an object at the position of {\object M31N 2008-12a} allows us to constrain the 2010 eruption time of this system.  The peak $R$-band detection of the 2014 eruption of \object{M31N 2008-12a} was at $m_{R}=18.2\pm0.1$, 0.038~days after the peak luminosity, which is consistent with the 2010 detection (within $2\sigma$). Because of the unfiltered nature of these observations, there are likely to be some systematic uncertainties in these magnitudes, due in some part to the strong array of emission lines observed in the nova spectrum, and also to the spectral energy distribution which may peak beyond the near-UV \citep{2015RN}.  If we assume that the Nov 20.503 detection corresponds to maximum light then, using the later upper-limits, we can constrain the decline time to be $t_{2}\leq6$ days, consistent with the 2014 eruption of \object{M31N 2008-12a} \citep[$t_{2}(R)=2.40\pm0.51$ days;][]{2015RN}.  The brightness of these detections, the positional coincidence, and the decline time limit, provide extremely compelling evidence that we have indeed recovered the elusive 2010 eruption of \object{M31N 2008-12a}.

\begin{figure}
\includegraphics[width=\columnwidth]{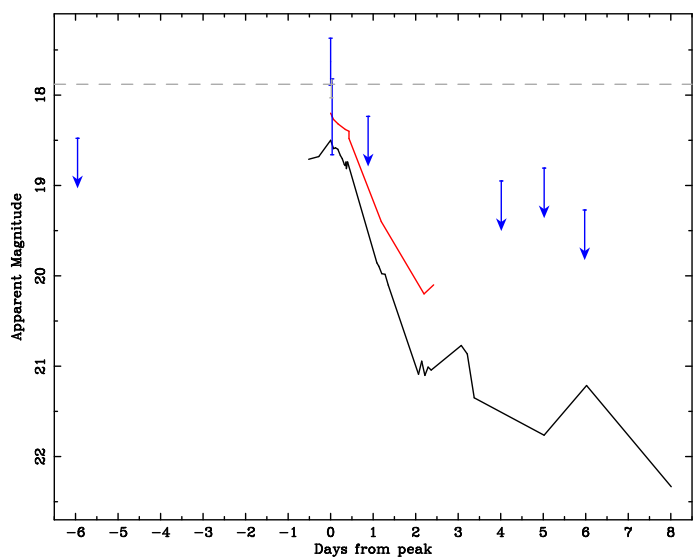}
\caption{\object{M31N 2008-12a} eruption template light curves based on $V$-band (black line) and $R$-band (red line) photometry from the 2008, 2011, 2012, 2013, and 2014 eruptions \citep{2014A&A...563L...9D,2015RN}.  The blue points indicate photometry of the 2010 eruption data (here we have assumed that the Nov 20.503 observation occurred at maximum light), the arrow heads indicate the $3\sigma$ upper limits from the non-detections, the tails the $5\sigma$ limits.  The grey dashed-line indicates the reference photometry of the nearby star \object{J004528.55+415451.7}.\label{fig1}}
\end{figure}

By combining the eruption photometric data published in \citet{2014A&A...563L...9D,2015RN}, we can update the template light curve of the eruption \citep[see their Fig.~1]{2015RN}.  The updated $V$- and $R$-band template light curves are plotted in Fig.~\ref{fig1} (black and red lines, respectively).  While fairly unconstrained, this Figure indicates that the 2010 detections fortuitously occurred extremely close to maximum light.

\subsection{Eruption statistics and evidence for a six-month period}
\label{sec:disc_stat}

\begin{figure}
  \includegraphics[width=\columnwidth]{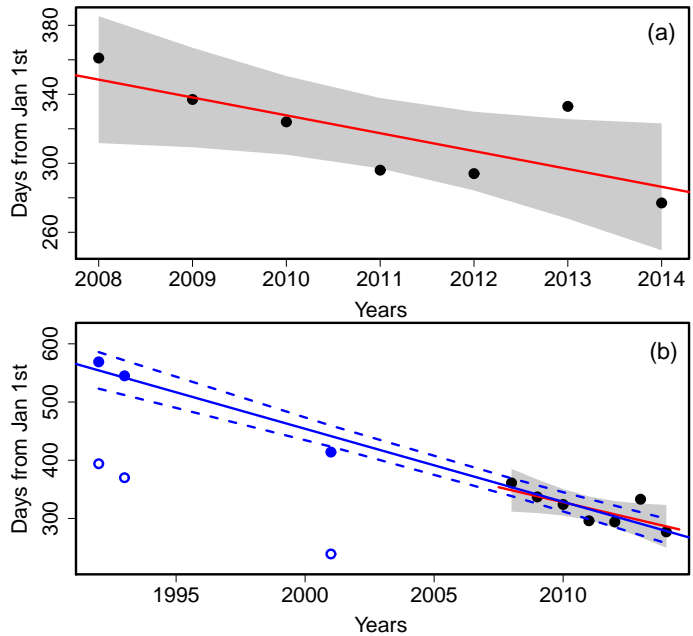}
  \caption{Distribution of eruption dates (in days of the year) over time, based on Table\,\ref{eruption_history}, for the optical detections only (panel \textbf{a}) and for all detections (panel \textbf{b}). The red line is the best fit to the (black) optical data. The grey area is the corresponding 95\% confidence region. In panel \textbf{b}) the blue open circles are the optical peak dates corresponding to the X-ray detections (see the text). The blue filled circles are the same dates shifted by 175~d. The blue solid line is the best fit to all solid data points and the blue dashed lines mark its 95\% confidence region.}
  \label{fig:stat}
\end{figure}

Now, with a series of seven eruptions in seven years we can begin to explore the statistics of the eruption time scales. The mean inter-eruption time is $351\pm13$\,d ($0.962\pm0.036$\,yr). Here, the $1\sigma$ uncertainty indicates the standard error of the mean. While this is still consistent with a 1~yr recurrence period it suggests that the actual period could be slightly shorter. The distribution of eruption dates over time is shown in Fig.\,\ref{fig:stat}a. Here, we plot the days of the year of the optical peaks (compare Table\,\ref{eruption_history}). The fitted model suggests with 99\% significance that successive eruptions are occuring at progressively earlier times during the year. Plotting the individual recurrence times over time does not reveal any trends.

To further investigate the possibility of a recurrence time slightly below 1~yr we revisited the historical X-ray detections listed in Table\,\ref{eruption_history}. Both ROSAT detections (in 1992/3) have good coverage of the emerging X-ray emission \citep{1995ApJ...445L.125W}. Therefore, we could use the well observed 2014 X-ray light curve \citep{2015Hen} as a template to extrapolate the dates of the optical peak with an accuracy of less than a day. The 2001 eruption was only detected in a single \textit{Chandra} observation \citep{2004ApJ...609..735W}. However, the short duration of the X-ray phase allowed us to constrain the optical peak to within two weeks, which is the size of the symbols in Fig.\,\ref{fig:stat}b.

In Fig.\,\ref{fig:stat}b we add the resulting dates as blue, open circles to the optical data. The two early dates were counted from the start of the previous year. There is little agreement with the trend found in Fig.\,\ref{fig:stat}a. This could of course indicate that no trend exists and that the random variance is high. However, if the three eruption dates are shifted upward by 175 days, which is half the apparent eruption period, we get an excellent agreement with the trend from Fig.\,\ref{fig:stat}a. The overall fit is significant at the $5\sigma$ level.

This finding is the first evidence that \object{M31N 2008-12a} might have in fact a recurrence period of $\sim6$ months. During the last seven years, the first eruption of each year would have occured while \m31 was too close to the sun to be observable. This situation is changing, evidenced by the progressively earlier second eruptions. If the period really is $\sim6$ months, then two eruptions per year should become observable soon: the overall fit predicts the first eruption in 2017 to occur around early March, with $1\sigma$ uncertainties of 24~d. Our ongoing quiescent monitoring of the system will test this prediction. It also allows us to confidently rule out even shorter periods.

We estimate the actual recurrence period to be $175\pm11$\,d ($0.48\pm0.03$\,yr). There is no evidence that this period has changed over the last $\sim20$~yr. The estimate takes into account the time between the 1992/3 eruptions and is consistent with the linear fit in Fig.\,\ref{fig:stat}b. Based on this fit we predict the next observable eruption to occur in late Sep 2015, with an $1\sigma$ prediction range from early Sep to mid Oct. We encourage additional observations.

\section{Conclusions}\label{conc}

In this Letter we have outlined the recovery of the elusive 2010 eruption of \object{M31N 2008-12a} and its impact on our knowledge of this exceptional object. Here we summarise our findings.

\begin{enumerate}
\item
We have uncovered the previously missing 2010 eruption of the remarkable recurrent nova \object{M31N 2008-12a}, from a pair of observations taken at 2010 Nov 20.52 UT. The photometry of the nova at this time $m=17.84\pm0.19$ is consistent with the nova being at or around maximum light.
\item
The sequence of seven eruptions in seven years strongly suggests that eruptions occur progressively earlier every year. We revisited archival X-ray data and found the resulting eruption dates to be consistent with this picture
\item
The dates of the X-ray detections provide the first evidence that \object{M31N 2008-12a} might have two eruptions per year. The corresponding average recurrence period is $175\pm11$\,d ($0.48\pm0.03$\,yr) and shows no evidence for change over a $\sim20$~yr time span. In this scenario, two eruptions per year should become observable soon. We predict the next eruption will occur in late Sep 2015.
\end{enumerate}

\begin{acknowledgements}
MH acknowledges the support of the Spanish Ministry of Economy and Competitiveness (MINECO) under the grant FDPI-2013-16933.
\end{acknowledgements}

\end{document}